%% file: main.tex
\documentclass[journal]{IEEEtran}

\usepackage{cite}
\ifCLASSINFOpdf
\else
\fi

\usepackage{amsmath,amssymb}
\usepackage{color}
\usepackage[caption=false,font=footnotesize]{subfig}
\usepackage{lipsum,booktabs,multirow}
\usepackage{url}
\usepackage{amsmath}
\usepackage{graphicx}
\usepackage{arydshln}
\usepackage[pdfencoding=auto, psdextra]{hyperref}

\makeatletter
\def\input@path{{./tables_final/}}
\makeatother
\graphicspath{{./figures_final/}}

\usepackage{accents}

\input{defined_commands}

\begin{document}

\title{Power Converter DC Link Ripple and\\Network Unbalance as Active Constraints in\\Distribution System Optimal Power Flow}

\author{Matthew~Deakin,~\IEEEmembership{Member,~IEEE,}
Rahmat~Heidari,
and~Xu~Deng,~\IEEEmembership{Member,~IEEE}
\thanks{M. Deakin and X. Deng are with Newcastle University, Newcastle-upon-Tyne, UK. M. Deakin was supported by the Royal Academy of Engineering under the Research Fellowship programme. X. Deng was
supported by Newcastle University Academic Track (NUAcT)
Fellowship scheme. R. Heidari was with the Commonwealth Scientific and Industrial Research Organization (CSIRO) and is with University of Queensland, Australia. Supported by NGED Open Data. Contact: \texttt{matthew.deakin@newcastle.ac.uk}.}
}

\maketitle

\begin{abstract}

The mitigation of unbalanced grid voltages or currents by voltage source converters results in power ripple on the dc link, and is a key converter design parameter due to hardware or stability considerations. Despite the importance of this issue for system design and operation, the use of Optimal Power Flow (OPF)-based methods capturing the interaction between dc link ripple and converter unbalanced operation has been largely unexplored. In this work, the magnitude of the power ripple is derived for generic multi-terminal converters, then introduced as a bilinear OPF constraint for two-level converter topologies. OPF case studies demonstrate the necessity to model both neutral current and dc link ripple, with tradeoffs between capacitor sizing and leg sizing highlighted for phase current unbalance mitigation applications. Time domain simulations of a grid-connected four-wire voltage source converter verify the accuracy and validity of the algebraic formulation. It is concluded that awareness of dc link ripple impacts and constraints will be of growing importance for distribution system operators.
\end{abstract}

\begin{IEEEkeywords}
Voltage source converter, dc link ripple, phase unbalance mitigation, phase rebalancer, voltage unbalance
\end{IEEEkeywords}

\IEEEpeerreviewmaketitle

\section{Introduction}

\IEEEPARstart{T}{he} number of proposed use-cases of Optimal Power Flow (OPF)-based techniques in electrical distribution networks has grown dramatically in the past two decades. These methods are made possible because decarbonization and electrification efforts that have resulted in an increasingly controllable, active distribution grid, replacing the passive, `fit-and-forget' grid of the 20th century. Power converters, based on fast switching of semiconductor devices, are an integral part of this system providing both an interface between generation, load, and storage assets and the grid, or as a means of controlling network powers to provide network capacity. As a result, there is increasing interest in developing appropriate approaches to model converters' operating costs and capability charts to enable full exploitation of these assets within OPF frameworks. Aspects considered include impact of network voltage on power injection limits; constraints on injections to dq control in three-phase inverters \cite{heidari2024improved}; or differences between grid forming versus grid following control \cite{li2023power}.

Distribution systems cannot typically be considered to have balanced currents and voltages. If there is a significant degree of unbalance, this needs to be taken into account by both distribution system OPF methods \cite{hong2016centralized}, and in power converter design and control. For unbalanced OPF problems, this is because voltage and current constraints are often binding on a per-phase basis, due to safety requirements on voltage magnitudes at single phase customers, and because current limits from branch elements and associate protection devices are most relevant per-phase. For multiphase power converters, unbalanced voltage or current injections leads to power ripple at twice the line frequency (hereon referred to as $\twoO $ ripple) on the dc link due to time-varying instantaneous power (see, e.g., \cite{akagi2017instantaneous}), leading to $\twoO$ voltage and current ripple. This $\twoO$ ripple is associated with increased losses and currents in the dc link capacitor \cite{van2018critical}, as well as a range of negative impacts on the systems connected onto the dc link, e.g., increased battery storage aging \cite{stecca2022battery} or torque ripple within motor drives \cite{pena2016dc}.

Despite the importance of $\twoO $ ripple on power converters, there are very few works that consider this as a component of the optimal operation of converters within a distribution network. Within the context of OPFs, the most common related constraint is to simply add a negative sequence voltage magnitude constraint \cite{antic2023importance}, although the physical justification as impacting on $\twoO $ ripple is not typically explicitly considered. In contrast, there are control-based approaches that have been used to track negative sequence voltages and provided compensation currents \cite{nejabatkhah2015control}. On the other hand, there has also been several works that study approaches for sharing of negative sequence currents in inverter-based microgrids \cite{castilla2025negative}. However, these latter methods are based on lower-level control, and are not able to exploit the `tertiary' level control of OPF problems, which are well-suited to consider network-wide tradeoffs as grid priorities evolve in time. In summary, to the authors' knowledge, there are no papers which explore which use OPF for scheduling of power converters in distribution grids which captures the interlink between $\twoO$ ripple and unbalanced converter operation. Given fast-growing interest in more effective control of power converters, this gap is both significant and timely.

The contribution of this work is to address this gap by proposing a mathematical formulation to explicitly include $\twoO $ power converter ripple within OPF problems that feature multiphase power converters operating with non-negligible unbalance. The formulation represents $\twoO $ ripple as a bilinear function of voltage and current phasors at the converter terminals. As compared to \cite{ziyat2024converter}, we incorporate these constraints into a network-wide OPF and explore the approach for a generic unbalanced injection (enabling the consideration of, e.g., back-to-back Soft Open Points). The use of the OPF framework also extends the tradeoffs studied in dynamic control design in \cite{roscoe2011tradeoffs} to more general cases, considering a wider area of the grid than only the point of common coupling. We also contrast our work with approaches that develop control strategies that assume a stiff grid \cite{yang2024coordinated}, with our OPF capturing interactions between dc link ripple and system unbalance.

The paper is structured as follows. In Section~\ref{s:operating}, we introduce two-level multi-terminal voltage source converters (VSCs) and derive the proposed $\twoO$ ripple constraints for these VSCs, illustrating the physical basis behind $\twoO $ power, voltage and current ripple on the dc link. We then demonstrate how the proposed formulation can be integrated into a distribution system OPF in Section~\ref{s:opf_integration}, explicitly linking unbalance to network constraints and potential objective functions. Section~\ref{s:results} demonstrates operation and design tradeoffs for a pair of complementary case studies, then conducts time-domain simulations to demonstrate the validity of the proposed ripple calculation. The future outlook is discussed in Section~\ref{s:discussion}, before salient conclusions are drawn in Section~\ref{s:conclusions}.

\emph{Notation:} For physical variables, lower-case symbols are used for time-domain quantities, whilst upper case symbols are used for the corresponding phasor representations. Superscript $\mathrm{dc}$ and $\twoO$ represent time domain or phasor components at dc or $\twoO$ frequency; subscript $\mathrm{dc}$ and $j$ refer to quantities associated with the dc link and phase $j$ respectively. Zero, positive, and negative sequence components are denoted by superscripts $\zro{(\cdot)},\,\pstv{(\cdot)},\,\ngtv{(\cdot)}$ respectively.

\section{Operational Constraints of Power Converter Systems under Unbalanced Operation}\label{s:operating}

\begin{figure}
\centering
\includegraphics[width=1\linewidth]{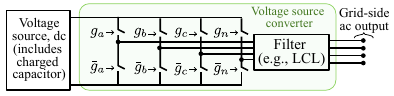}
\caption{Schematic of a 4-wire, 2-level VSC indicating the switching signals $g_j$ and their complement $\bar{g}_j$ used to generate ac voltages (and therefore ac currents) to inject power into the grid.}
\label{f:4w_vsc}
\end{figure}

The vast majority of active power converters in low- and medium-voltage applications have a VSC topology, and of those, the majority of converters are two-level topologies (particularly for low voltage distribution) \cite{schweizer2012comparative}. In this topology, the converter output connects to the distribution grid through an inductive (or LCL) filter, while its legs connect the converter side of the filter to connect to voltage sources formed by pre-charged large dc capacitors. By modulating the connection to these voltage points, the converter-side voltage can be controlled to produce an approximately sinusoidal current injection (see, e.g., \cite{mohan2012electric}). The output filter is designed to attenuate high frequency ripple components. A component-level schematic of a two-level four-leg VSC is shown in Fig.~\ref{f:4w_vsc}, showing the switching signals $g_j$ and their complement $\bar{g}_j$.

\begin{figure}\centering
\subfloat[Standalone 4W VSC (STATCOM)]{\includegraphics[height=5.0em]{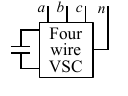}\label{f:vsc_statcom}}
~
\subfloat[B2B VSC (Soft Open Point)]{\includegraphics[height=5.0em]{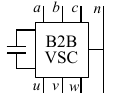}\label{f:vsc_sop}}
~
\subfloat[Interconnected 4W VSC]{\includegraphics[height=5.0em]{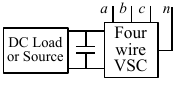}\label{f:vsc_interconnected_4w_vsc}}
\caption{Three examples of VSC-based system topologies.}
\label{f:vsc_varieties}
\end{figure}

Three exemplar VSC topologies that can provide unbalance mitigation are shown in Fig.~\ref{f:vsc_varieties}: a standalone four-wire (4W) VSC (i.e., without a dc-side load or source), hereon referred to as a STATCOM; a standalone back-to-back (B2B) VSC with common neutral, hereon referred to as a Soft Open Point; and an interconnected 4W~VSC with dc-side active power source. Additional VSC topologies may also be considered within this framework, such as interconnected B2B~VSC systems with dc-side source, B2B systems with mixed three- and single-phase connections, or reconfigurable VSCs \cite{cui2023two}.

In this section, we propose that considering only the magnitude of injected currents and power balance, as considered in prior works, is generally insufficient to capture the operational constraints of VSCs operating under unbalance (i.e., that $\twoO$ power ripple should be considered as a constraint). We first define these operational constraints mathematically, demonstrating how a bilinear function of VSC terminal voltages and currents define the $\twoO$ dc link power ripple, showing that this closely links to $\twoO$ current and voltage ripple in the dc link capacitor (Section~\ref{ss:op_contraints}). Example operational and design tradeoffs for a STATCOM and Soft Open Point then highlight implications of these constraints (Section~\ref{ss:decomposition}).

\subsection{Operational Constraints for a Converter Injecting Generic Unbalanced Currents}\label{ss:op_contraints}

\begin{figure}
\centering
\includegraphics[width=0.4\textwidth]{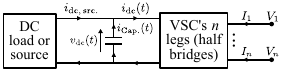}
\caption{The half-bridges of a generic VSC with $ n $ legs (not including filter).}
\label{f:generic_vsc}
\end{figure}

Fig.~\ref{f:generic_vsc} shows a generic VSC half-bridge stage with $n$ legs, with the ac fundamental voltage and current represented by phasors $V_{j},\,I_{j}$ for the $j$th leg. The leg current capacity constraint and VSC current balance equation are denoted
\begin{equation}\label{e:iphase}
|I_{j}| \leq I_{j}^{\max} \,,
\end{equation}
\begin{equation}\label{e:i_bal}
\sum _{j} I_{j} = 0\,.
\end{equation}

We derive the algebraic form of $\twoO$ power ripple by considering instantaneous power. The active power $p_j$ of leg $j$ from fundamental frequency voltage and current is
\begin{equation}\label{e:inst_p_mult}
p_{j}(t) = 2|V_{j}||I_{j}| \cos \left (\omega t + \thVi \right ) \cos \left (\omega t + \thIi \right ) \,,
\end{equation}
where $\theta^{(\cdot)}_{j}$ represents the phase angle of a phasor. Using suitable trigonometric identities, \eqref{e:inst_p_mult} can be rewritten,
\begin{equation}\label{e:inst_p_decomp}
p_{j}(t) = |V_{j}||I_{j}| \left ( \cos \left ( \thVi - \thIi \right ) + \cos \left (\twoO t + \thVi + \thIi\right ) \right )\,.
\end{equation}
To consider algebraically the summation of components of \eqref{e:inst_p_decomp} over multiple legs (to determine the dc link ripple), we first define a $\twoO$ active power ripple phasor,
\begin{equation}
    P_{j}^{\twoO} = V_j I_j\,,
\end{equation}
which defines an appropriate $\twoO$ phasor for \eqref{e:inst_p_decomp}) as
\begin{equation}\label{e:p2w_phasor}
    p_{j}^{\twoO}(t) = |P_{j}^{\twoO}| \cos \left ( \twoO + \angle P_j^{\twoO} \right)\,.
\end{equation}
We then define the familiar (dc) complex power, $S_{j}^{\mathrm{dc}}$,
\begin{equation}
S_{j}^{\mathrm{dc}} = V_{j}I_{j}^{*} = |V_{j}||I_{j}| \left( \cos \left ( \thVi - \thIi \right ) + \jmath \sin \left ( \thVi - \thIi \right ) \right) \,, \label{e:sdc_phasor}
\end{equation}
implying dc time-domain component $p_j^\mathrm{dc}(t)$ of \eqref{e:inst_p_decomp} as
\begin{equation}
p_j^{\mathrm{dc}}(t) = |V_{j}||I_{j}| \cos \left ( \thVi - \thIi \right ) \,. \label{e:pdc_phasor}
\end{equation}
Note that the behavior of \eqref{e:pdc_phasor} differs from \eqref{e:p2w_phasor} with respect to the magnitude of the active power components considering phase. If voltage and current are in quadrature, then the dc component of active power \eqref{e:pdc_phasor} will have a value of zero (hence, the phasor is represented using the apparent power symbol $ S $). In contrast, for non-zero $I_{j}$ and $V_{j}$, the magnitude of the oscillating $\twoO $ active power component for each leg is \textit{independent} of phase (hence, its phasor is represented using the active power symbol $ P $). 

The $\twoO$ phasor \eqref{e:p2w_phasor} enables the consideration of interference of active power ripple due to multiple legs. Active power balance across the ac and dc terminals of the VSC's half-bridges implies that
\begin{equation}\label{pdclink_pi}
p_{\mathrm{dc}}(t) = \sum _{j} p_{j}(t)\,.
\end{equation}
The dc link active power phasors in the dc and $\twoO$ components $ \PdcDC,\, \Pdcww$ respectively, are 
\begin{align}
\PdcDC &= \sum _{j} \mathrm{Re}(V_{j}I_{j}^{*})\,, \label{e:dc_p_bal} \\
\Pdcww &= \sum _{j} V_{j}I_{j}\,. \label{e:2w_p_bal}
\end{align}
The power ripple on the dc link therefore
\begin{equation}\label{e:p_ripple}
p_{\mathrm{dc}}^{\twoO}(t) = \left | \Pdcww \right |\cos \left (\twoO t + \angle (\Pdcww) \right )\,.
\end{equation}
A constraint on the magnitude of the $\twoO$ dc link ripple can therefore be formulated as a constraint
\begin{equation}\label{e:2w_ripple_cons}
\left |\Pdcww \right |\leq P_{\mathrm{dc}}^{2\omega,\,\max}\,.
\end{equation}
Equation \eqref{e:dc_p_bal} represents dc power balance and is well-known. In contrast, to our knowledge, \eqref{e:2w_ripple_cons} has not been studied in the context of distribution system OPF problems.

\emph{Remark: high-frequency dc link ripple.} There is also some high-frequency ripple on the dc link that is caused by the PWM switching action. However, this ripple is inversely proportional to switching frequency. The ripple induced at typical converter switching frequencies is therefore typically insignificant as compared to the $\twoO $ ripple \cite{pei2014analysis}. 

\subsubsection{$\twoO$ Ripple as a VSC Design and Operational Constraint}\label{ss:ripple_constraints}
Generally, there are several reasons a VSC may face operating constraints related to $\twoO $ ripple \eqref{e:2w_ripple_cons}. First, the widespread use of \emph{dq} control in converters \cite{lin2020research} restrict them to injecting only positive sequence currents (for a stationary \emph{dq} reference signal). Second, the dc link capacitor must have sufficient energy storage to absorb the power ripple from unbalanced injection without causing voltage ripple \cite{ziyat2023voltage}. Third, higher current ripple causes heating, which leads to thermal stress \cite{kolar2006analytical} and shortens capacitor lifetime \cite{wang2014reliability}. (Capacitor failure is a dominant failure mode in VSCs--for example, inverter lifetime is often assumed to match capacitor lifetime \cite{gandhi2018reactive}.) Finally, components connected directly to the dc link, such as batteries, may also experience increased thermal stress and subsequent degradation if their equivalent Thevenin impedance is similar (or lower) than that of the dc link capacitor at the ripple current frequency \cite{stecca2022battery}.

With regards to dc link capacitor design and operation, the impact of active power ripple on $\twoO$ voltage and currents can be derived, leading to constraints of the form of \eqref{e:2w_p_bal} analytically. \eqref{e:p_ripple} suggests a trial solution for dc link currents as
\begin{align}\label{e:vdc_trial}
v_{\mathrm{dc}}(t) &= \VdcO + |V_r|\sin \left ( \twoO t + \angle V_r \right )\,,\\
i_{\mathrm{dc}}(t) &= \IdcO + \underbrace{|I_{r}| \cos \left ( \twoO t + \angle V_r \right)}_{i_{\mathrm{cap.}}(t)}\,,\label{e:idc_trial}
\end{align}
where it is assumed that the dc source current has a low pass filter-type frequency response, so that $\IdcO$ does not change significantly at the $\twoO$ frequency. By considering \eqref{e:vdc_trial} and definition of capacitance, the phasor $I_r$ representing the capacitor ripple current $i_{\mathrm{cap.}}(t)$ is
\begin{equation}\label{e:ir_value}
I_{r} = \jmath V_{r}\twoO C\,.
\end{equation}
The dc link active power ripple is therefore
\begin{multline}\label{e:pdc_trial}
p_{\mathrm{dc}}(t) = 
\IdcO \left(\VdcO + |V_{r}|\sin ( \twoO t + \angle V_r ) \right)  \\ +
\VdcO |V_{r}|  \twoO C \cos \left ( \twoO t + \angle V_r \right ) \\ + |V_{r}|^{2} \twoO C \cos \left ( 4\omega t + 2(\angle V_r) \right )\,.
\end{multline}
We consider the approximation $\IdcO \ll \VdcO \twoO C$, as, on the output power basis of a VSC, the total ripple power has a per-unit value of 1 pu (by comparing \eqref{e:dc_p_bal}, \eqref{e:2w_p_bal}); a maximum typical permissible voltage ripple $V_r$ of 0.1~pu will only permit a small fraction of power ripple to be met by the term $\IdcO V_{r}$ (this approximation follows, e.g., \cite{brooks2023dc}). Therefore, equating $ \twoO $ terms\footnote{(Small) uncompensated quadruple line frequency components in the instantaneous active power will also balance, as the dc voltage ripple leads to some voltage ripple being introduced into the AC networks and therefore power ripple--as considered in, e.g., \cite{ziyat2023voltage}.} in \eqref{e:p_ripple} and \eqref{e:pdc_trial}, neglecting the $\sin$ term of \eqref{e:pdc_trial}, we can write down that $ \angle V_r=\angle (\Pdcww ) $ and
\begin{equation}\label{e:v_ripple}
|V_{r}| = \dfrac{\left | \Pdcww \right |}{\twoO C \VdcO}\,,\quad |I_{r}| = \dfrac{\left | \Pdcww \right |}{\VdcO}\,.
\end{equation}
Considering that the equivalent series resistance (ESR) for a given capacitor is inversely proportional to its capacitance, the capacitor losses $ P_{\mathrm{cap.\,losses}} $ are
\begin{equation}\label{e:cap_loading}
P_{\mathrm{cap.\,losses}}^{\twoO} \propto \dfrac{\left | \Pdcww \right |^{2}}{\VdcO^{2} C}\,.
\end{equation}
Considering \eqref{e:v_ripple} and \eqref{e:cap_loading}, if there are constraints on voltage ripple $|V_r|$, current ripple $|I_r|$, or thermal constraints due to losses $P_{\mathrm{cap.\,losses}}^{\twoO}$, each can be represented in the form \eqref{e:2w_ripple_cons}.

\subsection{Linking $\twoO$ Ripple to 4W and Back-to-Back VSC Unbalanced Operation}\label{ss:decomposition}

To study the impact of \eqref{e:2w_ripple_cons} on VSCs under unbalance, it is useful to transform $abc$ values into sequence components via the Fortescue transformation, with the symmetrical basis $ B_{\mathrm{Fort.}} $ for this transformation
\begin{equation}\label{e:fort_basis}
B_{\mathrm{Fort.}} 
= 
\left [
\begin{array}{ccc}
\hat{\zro{b}} & \hat{\pstv{b}} & \hat{\ngtv{b}} \\
\end{array}
\right ]
=
\left [
\begin{array}{ccc}
1 & 1     	& 1  \\
1 & \alpha^{-1}	    & \alpha^{-2} \\
1 & \alpha^{-2}     & \alpha^{-4} \\
\end{array}
\right ]\,,
\end{equation}
where $ \alpha=e^{\jmath 2\pi/3} $ is the 120$^{\circ}$ phase rotation operator, and the corresponding transformation matrix $\Tfort = B_\mathrm{Fort.}^{-1}$. Using this transformation, the equation for $\twoO$ ripple for 4W~VSCs in terms of sequence components can be determined \cite{akagi2017instantaneous}
\begin{align}\label{e:ripple_abcd}
\Pdcww &= \sum _{j\in\{a,b,c,n\}} V_{j}I_{j} \\
 &= \zro{V}\zro{I} + \pstv{V}\ngtv{I} + \ngtv{V}\pstv{I} \label{e:full_ripple_sequence} \\
 &\approx \pstv{V}\ngtv{I} \,, \label{e:ripple_sequence}
\end{align}
where \eqref{e:ripple_sequence} assumes that $|\pstv{V}| \gg |\ngtv{V}|$ and $|\pstv{V}| \gg |\zro{V}|$ (as is typically the case during normal operation).

\subsubsection{Tradeoff Between Neutral Current and $\twoO$ ripple}\label{sss:neutral_cap_tradeoff}

One objective for a STATCOM (Fig.~\ref{f:vsc_statcom}) may be to reduce the power being transferred on one specific phase, either for reducing peak currents or voltages, particularly when that current is much higher than the mean current in a multiphase element (e.g., a three-phase overhead line). It is therefore of interest to consider how \eqref{e:2w_ripple_cons} may restrict the maximum power that can be transferred to any one phase from the STATCOM.

Consider that the active power on phase $j=a$ is substantially higher than the mean current and the power factor is unity, then the power injected into phase $a$ should be in antiphase with the existing current in the line. In this case, the locus of current injections can be defined using a parameter $0 \leq \gamma \leq 1$ that satisfy active power balance, current balance, and that the phase $a$ current and voltage should be in phase (so the mitigation current onto phase $a$ is maximized), as,
\begin{equation}\label{e:neutral_vs_ripple}
    I = \gamma |I| \hat{\ngtv{b}} + (1-\gamma) |I| \hat{\zro{b}}\,.
\end{equation}
Considering \eqref{e:ripple_abcd}, \eqref{e:neutral_vs_ripple}, and neutral current $I_n$ as triple the zero sequence current $\zro{I} $ (by \eqref{e:fort_basis}), it can be noted:
\begin{itemize}
    \item A negative sequence current restricted to zero ($\gamma = 0$) will have no $\twoO$ ripple; however, the neutral current will be maximized (having value $3$ times the phase current).
    \item A zero sequence current restricted to zero ($\gamma = 1$) has zero neutral current (i.e., a three-wire converter can inject the phase power), but will maximize the ripple power (having value equal to the VSC's kVA rating, by \eqref{e:2w_p_bal}).
    \item For $\gamma=2/3$, the neutral current is the same current as the phase currents. Therefore, for a neutral leg with the same capacity as phase legs $a,\,b,\,c$, the unbalance mitigation that can be provided will only be reduced if $\twoO$ ripple is constrained to be less than $2/3$ of the active power rating of the VSC.
\end{itemize}
It is worth noting that restricting the dc link ripple power or the neutral current in isolation may not affect the maximum per-phase current that can be injected; only when they are \emph{both} restricted will a change in VSC operation be clearly detected. Case studies explore this in Section~\ref{ss:res_statcom_case}.

\subsubsection{Ripple Cancellation}\label{sss:sop_zero_ripple}
It is also of interest to study cases where there are more than four legs, such as the Soft Open Point (Fig.~\ref{f:vsc_sop}). For such a system, injecting a non-zero negative sequence current need not necessarily result in a current ripple \eqref{e:ripple_sequence}. This is because, if an appropriate negative sequence current is injected onto the second feeder to which the Soft Open Point is connected, then the ripple currents will cancel (in the same way that $\twoO$ ripple powers cancel for a three-phase VSC injecting balanced currents). This operation is explored in more detail in Section~\ref{ss:res_sop_case}.

\section{Associating $\twoO $ Dc Link Ripple with Distribution System OPF Problems}\label{s:opf_integration}

In the previous section, we have demonstrated that $\twoO $ ripple is a bilinear function of the current injection and voltage at the VSC's half bridge terminals, and presented several physical and operational motivations for reducing this ripple for the VSC. In practice, however, $\twoO$ ripple mitigation must be balanced against other network objectives, motivating the use of OPF. In this Section, we demonstrate this link by first considering voltage unbalance constraints, power quality constraints, and their link to customer-side motor performance degradation (Section~\ref{ss:unbalance_in_opf}). We then conclude the methodology by formally introducing the proposed extensions to an appropriate distribution OPF formulation (Section~\ref{ss:opf_optimization}).

\subsection{Voltage unbalance within OPF objectives and constraints}\label{ss:unbalance_in_opf}

Within a three-phase power distribution network, voltage unbalance constraints are imposed to ensure proper operation of three-phase loads. Assuming normal grid operations, such that $\pstv{V}\approx \hat{\pstv{b}}$ (in pu), three-phase loads and generators will operate correctly as long as the negative sequence voltage unbalance limit ${\ngtv{V}}^{\max}$ is satisfied \cite{girigoudar2019relationships}, i.e.,
\begin{equation}
\label{e:Vneg_limit}
|\ngtv{V}_b| \leq {\ngtv{V}}^{\max}\,,
\end{equation}
where $\ngtv{V}_b$ is determined through the Fortescue transformation \eqref{e:fort_basis} at bus $b$, and the unbalance limit is typically set as 2\% \cite{girigoudar2019relationships}. The constraint \eqref{e:Vneg_limit} should be omitted if no three-phase loads or generators are connected at the bus.

\begin{figure}
\centering
\includegraphics[width=0.37\textwidth]{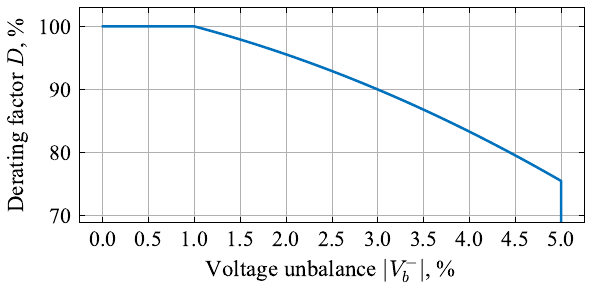}
\caption{Induction motor derating factor as a function of voltage unbalance.}
\label{f:derating}
\end{figure}

Some loads may also experience degraded performance due to voltage unbalance. The most well-known example is direct-connected induction motors. In this work, we account for the required derating of induction motors $D_b$ at bus $b$ (in \%) according to a curve inspired by the IEC standard 60034-26 derating curve (with $\ngtv{V}_b$ here represented in pu),
\begin{equation}\label{e:derating_factor}
D_b(\ngtv{V}_b) = 
\begin{cases}
100 & \mathrm{if} \; |\ngtv{V}_b| < 0.01 \\
100 - g(\ngtv{V}_b) & \mathrm{if} \; 0.01 \leq |\ngtv{V}_b | < 0.05 \\
0 & \mathrm{otherwise\,.}
\end{cases}
\end{equation}
The exact form of $g$ is not given in \cite{bsi2006rotating} and so it is not possible to use the IEC standard directly. The voltage-unbalance derating factor is therefore based on the following quadratic function,
\begin{equation}\label{e:derating_func}
    g(\ngtv{V}_b) = a_2 |\ngtv{V}_b|^2 + a_1 |\ngtv{V}_b| + a_0\,,
\end{equation}
with coefficients $a_0=0.033125\%,\,a_1=2.75\% \mathrm{pu}^{-1},\,a_2=56.25\% \mathrm{pu}^{-2}$ resulting in a close fit to the values presented in \cite{bsi2006rotating}. The resulting derating function is shown in Fig.~\ref{f:derating}.

Based on the derating $D$, induction machine rating $S_b^{\mathrm{IM}}$ at bus $b$ and a uniform cost associated with derating, the network-wide cost of induction machine derating $C^{\mathrm{Derating}}$ is then
\begin{equation}\label{e:cost_derating}
C^{\mathrm{Derating}} = w^{\mathrm{Derating}} \sum_{b\in B^{\mathrm{IM}}} S_b^{\mathrm{IM}} \left ( 100 - D_b(\ngtv{V}_b) \right )\,,
\end{equation}
where $B^{\mathrm{IM}}$ is the set of buses with three-phase directly connected induction machines, and $w^{\mathrm{Derating}}$ is an appropriate weighting factor (e.g., \$/kW).

\subsection{Unbalanced OPF Considering DC Link Ripple}\label{ss:opf_optimization}

The OPF problem defined by the previous sections are carried out in the current-voltage rectangular with explicit neutral (`IVR-EN') formulation described in \cite{CLAEYS2022108522} using PowerModelsDistribution.jl \cite{fobes2020powermodelsdistribution}, integrating the unbalanced converter modeling extensions developed in \cite{heidari2024improved} and the model developments proposed in this paper. Table~\ref{table_OPF_IVR} shows the feasible region of VSC converter models developed in this paper (see Section~\ref{ss:res_statcom_case} and Section~\ref{ss:res_sop_case} for definitions of the objective functions \eqref{e:oF1} and \eqref{e:oF2} respectively).

\begin{table}
\label{table_OPF_IVR}
\input{table_OPF_IVR}
\end{table}

\section{Results}\label{s:results}

\begin{figure*}
\centering
\includegraphics[width=0.8\textwidth]{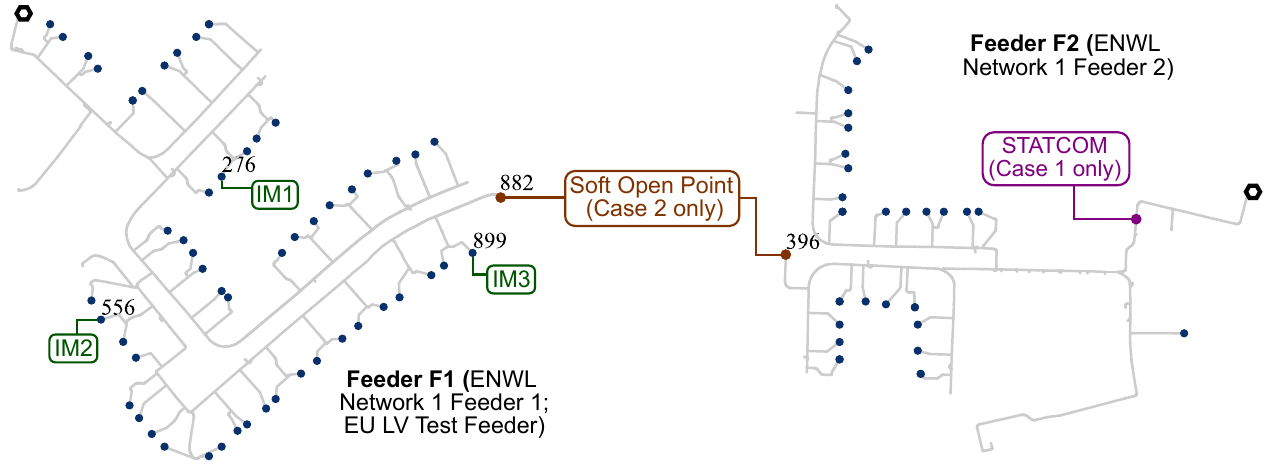}
\caption{Case studies considered feeders F1 and F2 \cite{enwl2014lvns}. Cases 1a--1f considers phase current balancing to mitigate overloads through a STATCOM on F2; Case 2 considers mitigating induction machine (IM) derating due to voltage unbalance via a Soft Open Point providing interconnection between F1 and F2.}
\label{f:network_case_studies}
\end{figure*}

In this section, we study the interactions between VSC $\twoO$ ripple and unbalance in the context of two IV-REN OPF problems, to show how the proposed algebraic constraints can impact network operations, using a STATCOM (Section~\ref{ss:res_statcom_case}) and a Soft Open Point (Section~\ref{ss:res_sop_case}). These network case studies consider networks from Electricity Northwest Ltd (ENWL), including the IEEE EU LV test feeder (see Fig.~\ref{f:network_case_studies}). Finally, we study a time-domain simulation of a two-level VSC to validate of the proposed $\twoO$ ripple calculation (Section~\ref{ss:res_ripple_validation}). Implementations are available from \cite{deakin2025models}.

\subsection{Case A: $\twoO $ Ripple and Current Unbalance for a STATCOM}\label{ss:res_statcom_case}

\begin{figure}
\centering
\includegraphics[width=0.48\textwidth]{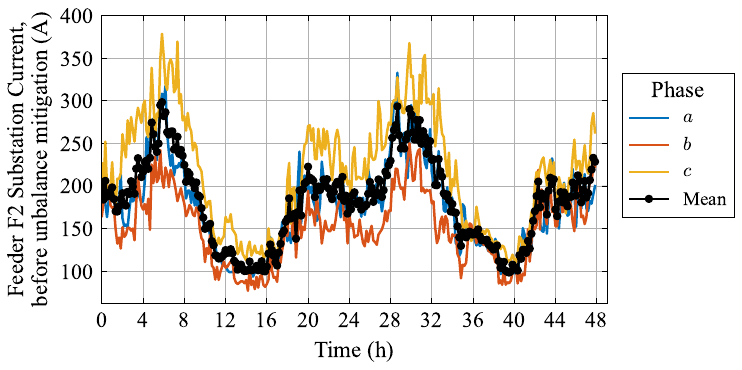}
\caption{Substation current for Cases 1a--1d before the 20.7~kVA STATCOM is introduced, showing the potential for a large reduction in peak capacity for significant time periods via phase balancing.}
\label{f:base_demand}
\end{figure}

In the first case, we consider the impact of the operation of a STATCOM for mitigating current unbalance. Two months of half-hourly data has been collected from an LV substation monitor \cite{nged2025demand} (transformer ID: 720339, start date: 22/5/2025) to consider as the distribution substation demand. Fig.~\ref{f:base_demand} plots the measured and mean phase current for the first 48 hours, showing that the peak current in the substation is more than 60~A greater than the mean; mitigating this could reduce loading on phase $c$ by up to 20\%. It is therefore possible to increase the effective capacity of the three-phase system by injecting power from other phases onto the phase with higher current to reduce the peak via a STATCOM.

To address this issue, we consider a STATCOM (Fig.~\ref{f:vsc_statcom}) with no dc source and with 30~A legs on phases $a,\,b,\,c$ (i.e., rated power of 20.7~kVA). The objective function for the proposed OPF (Section~\ref{ss:opf_optimization}) is the minimization of the maximum phase current,
\begin{equation}\label{e:oF1}
\mathrm{OF}_1 = \max I_j, \quad \text{for } j \in \{a,\,b,\,c,\,n\}\,.
\end{equation}
To consider the tradeoff between the neutral current and $\twoO$ ripple for minimizing this objective (Section \ref{sss:neutral_cap_tradeoff}), we study four subcases, each constraining the maximum $\twoO$ ripple or neutral current, as summarized in Table~\ref{t:case1}. Case 1a considers operation with unconstrained neutral current and $\twoO $ dc link ripple unconstrained. Cases 1b and 1c fully restrict to zero the neutral or $\twoO$ ripple power, respectively (the former representing a three-wire VSC topology). Case 1d explores a case with neutral leg sized the same as the phase legs (30~A), and $\twoO$ ripple restricted to $\approx 1/4$ of the VSC's total kVA rating (and less than the $2/3$ minimum requirement for unconstrained operation, as highlighted in Section~\ref{ss:decomposition}).
\begin{table}
\input{case1}
\label{t:case1}
\end{table}

\begin{figure}
\centering
\includegraphics[width=0.44\textwidth]{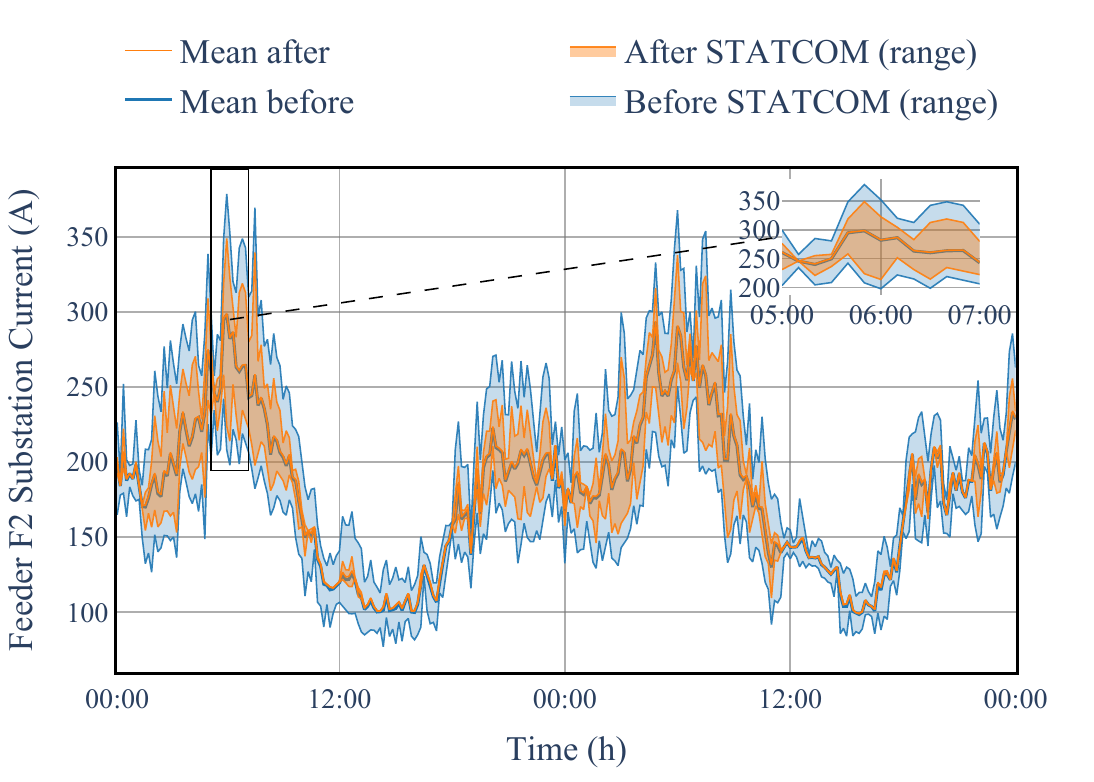}
\caption{The substation range of phase currents before and after the application of a STATCOM for Case~1a, showing the STATCOM effectively bringing the peak current 30~A closer to the mean (or exactly to the mean).}
\label{f:Iabc_source}
\end{figure}

Results of the OPF for Case 1a are shown in Fig.~\ref{f:Iabc_source}. The shaded blue area depicts peak-to-peak and mean phase current magnitudes before the STATCOM is installed, whilst the peak-to-peak phase current magnitude with STATCOM is shown by the shared orange area. It can be seen by-eye from this figure that the range is reduced by up to 30~A across the 48 hours, thereby being fully utilized to address the peak currents.

\begin{figure}
\centering
\includegraphics[width=0.44\textwidth]{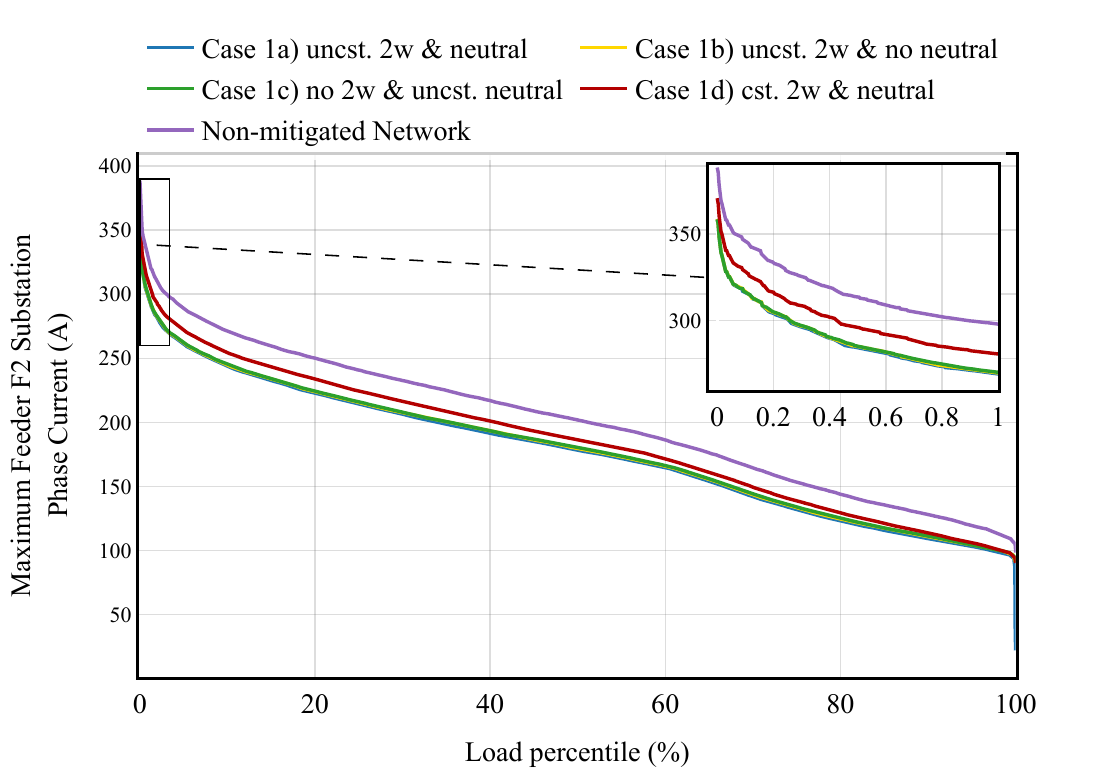}
\caption{Load duration curves of the maximum phase current at F2's substation over the two months, showing Cases 1a-1c effectively reducing the peak phase current by 30~A, whilst Case 1d reduces the peak current by only 20~A.}
\label{f:Imax_src}
\end{figure}

To demonstrate the effectiveness of the control across the full two-month period, load duration curves of the maximum phase current for Cases 1a--1d are shown in Fig.~\ref{f:Imax_src} against the uncompensated substation current (labeled as `Non-mitigated Network'). It can be seen that Cases 1a, 1b and 1c reduce the peak current by 30~A, whereas for Case~1d, the maximum phase current is reduced by only 20~A. This indicates that, from the perspective of effectiveness, all four approaches can provide between 20~A to 30~A unbalance compensation over a wide range of operation, with Case~1d having reduced effectiveness to provide unbalance mitigation as expected due to its reduced $\twoO$ ripple and neutral capabilities (Section~\ref{sss:neutral_cap_tradeoff}).

\begin{figure}
\centering
\includegraphics[width=0.44\textwidth]{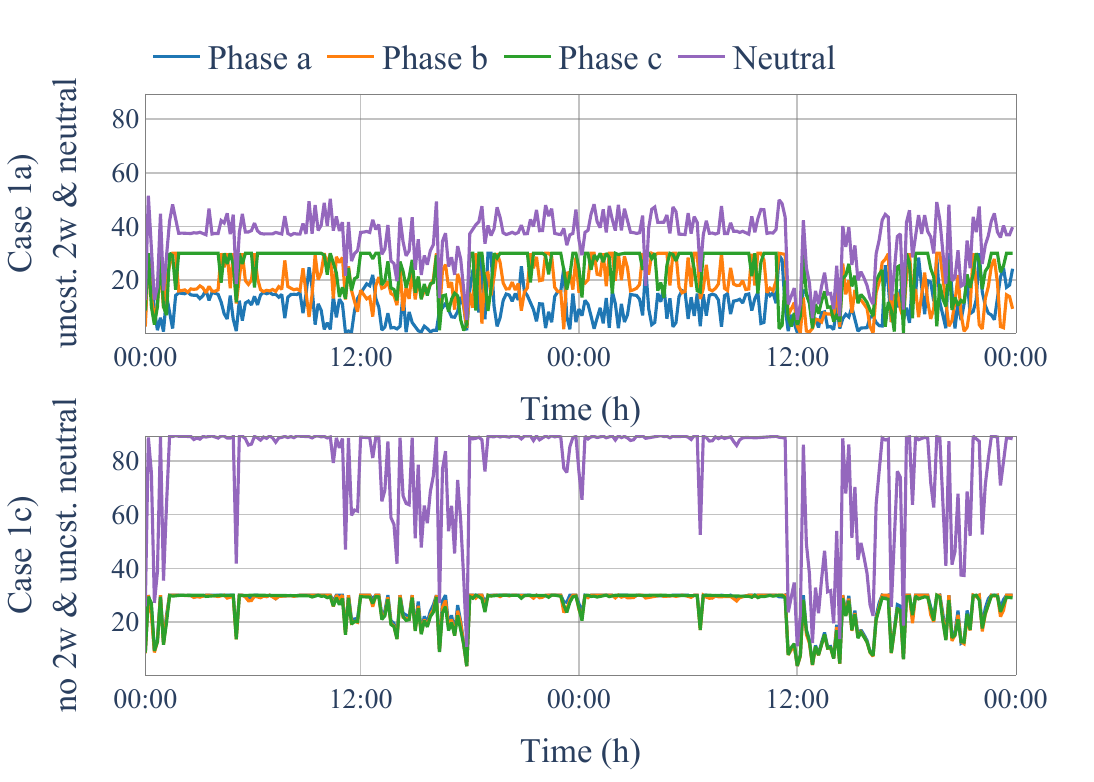}
\caption{Comparing the phase and neutral currents for Case~1a and Case~1c, showing the neutral current is often up to 50\% higher than the phase currents for Case~1a, but is triple the phase current for Case~1c.}
\label{f:Iabcn_timeseries}
\end{figure}

Although the phase mitigation capabilities of the four cases are similar, there is a significant difference between the device operation (in terms of current injections and $\twoO$ ripple) for each case. For instance, Fig.~\ref{f:Iabcn_timeseries} plots the phase and neutral current for Case~1a and Case~1c. Case~1c can only inject zero sequence current, so the phase currents of each leg are identical (and in phase), with the neutral current magnitude three times the phase current \eqref{e:i_bal}. As a result, although the active phase legs $a,\,b,\,c$ are required to pass 30~A, the neutral leg must meet 90~A of current (Fig.~\ref{f:Iabcn_timeseries}, lower), significantly higher than the current of Case~1a (Fig.~\ref{f:Iabcn_timeseries}, upper).

\begin{figure}
\centering
\includegraphics[width=0.44\textwidth, height=5cm]{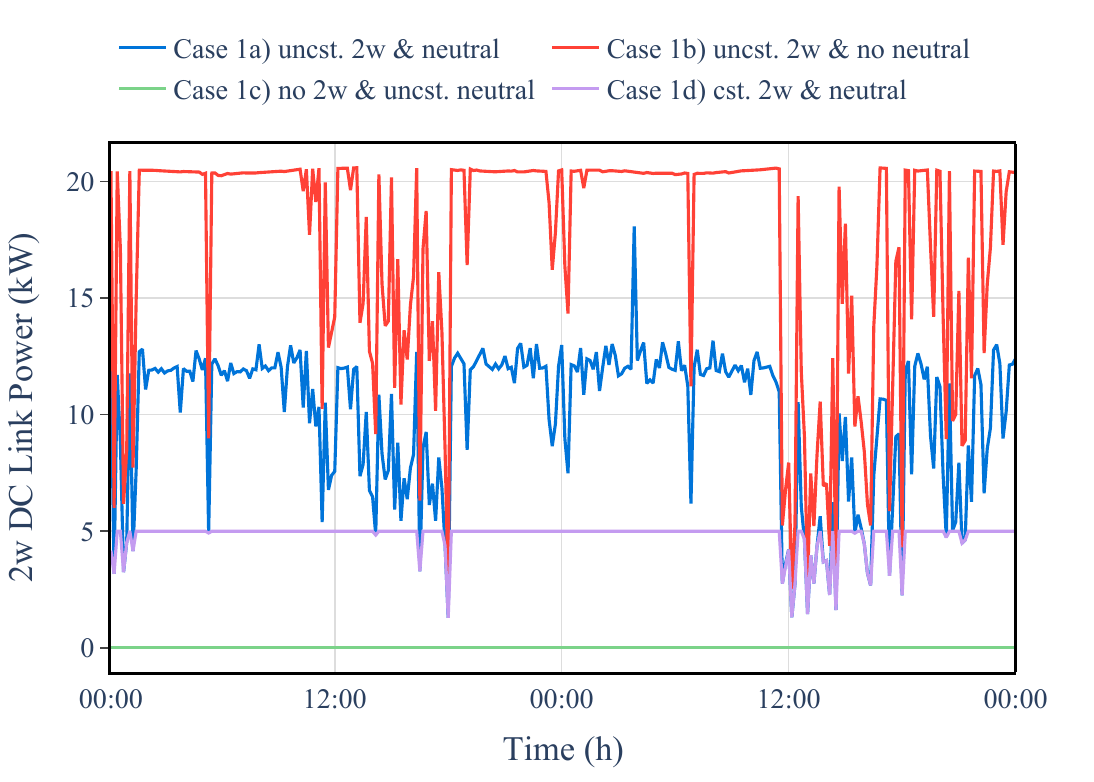}
\caption{$\twoO$ dc link power for Cases 1a--1d, showing that all cases follow constraints outlined in Table~\ref{t:case1}, and that the ripple power does not exceed the kVA rating of the device (20.7 kVA).}
\label{f:Pdclink_case1a}
\end{figure}

Similarly, there is a significant difference in $\twoO$ ripple power between the cases, as shown in Fig.~\ref{f:Pdclink_case1a}. For example, Case~1c, has a high neutral current, but no $\twoO$ ripple (as expected); on the other hand, Case~1b has no neutral current (as a three-wire converter), but requires full rated ripple power (20.7~kW). Case~1d, on the other hand, shows only 5~kW of ripple, as expected, although this comes at the aforementioned cost of reduced effectiveness at mitigating the phase unbalance (Fig.~\ref{f:Imax_src}), due to its neutral current restrictions.

\begin{figure}
\centering
\includegraphics[width=0.44\textwidth]{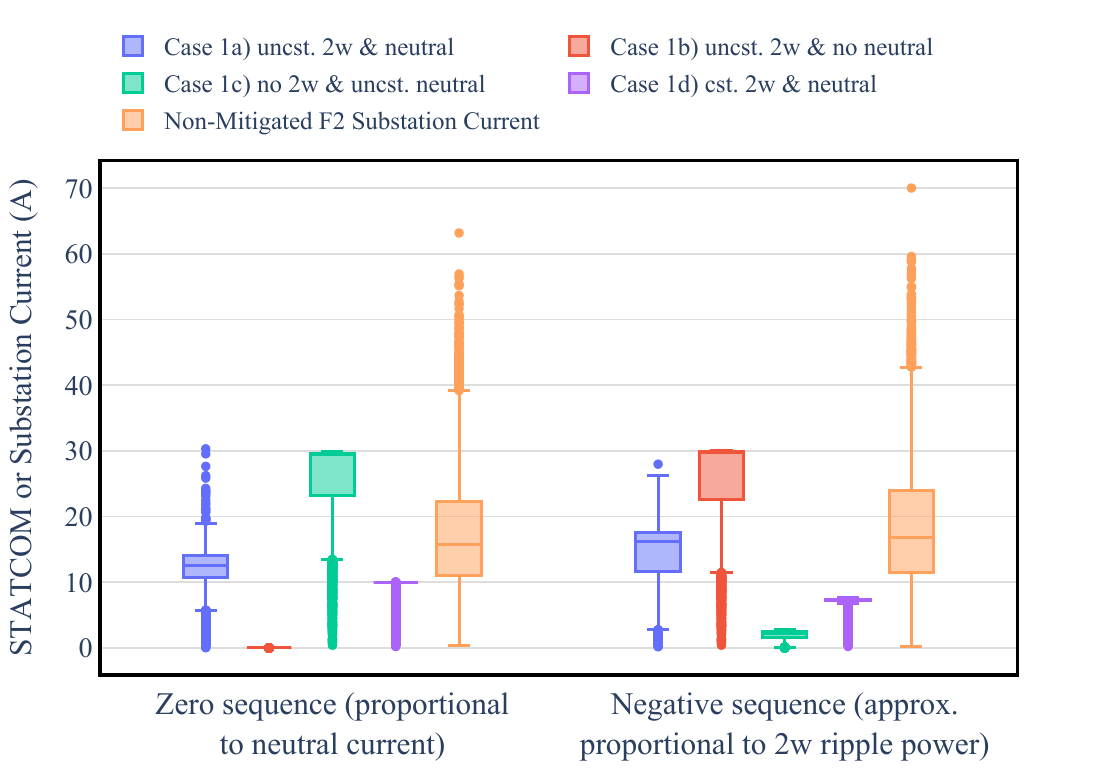}
\caption{Negative and zero sequence current injections for Cases 1a--1d for the STATCOM.}
\label{f:Iseqs_boxplot}
\end{figure}

Fig.~\ref{f:Iseqs_boxplot} summarizes these results by showing the zero- and negative-sequence components of the STATCOM current, referenced in comparison to the substation's unmitigated sequence currents. In Cases 1a and 1d, where some $\twoO$ ripple and neutral current are permitted simultaneously, the zero- and negative-sequence currents are nonzero and scale proportionally with their respective limits. For the three-wire case (Case 1b), the zero sequence current is zero; for the four-wire case with no $\twoO$ ripple (Case 1c), there no dc link ripple but only a small amount of negative sequence ripple \eqref{e:full_ripple_sequence} (there is non-zero $\ngtv{V}$ and $\zro{V}$ in this model).

\subsection{Case B: $\twoO $ Ripple and Voltage Unbalance for a Soft Open Point}\label{ss:res_sop_case}

To study a case whereby voltage unbalance must be mitigated, we consider the full system with both feeders F1 and F2 (Fig~\ref{f:network_case_studies}). We consider a nominal balanced source voltage of 1.05~pu for F2; and an unbalanced source voltage for F1 with negative sequence component of 3.18\% (with the exception of indicated IMs, there are no three-phase customers or devices requiring unbalance below 2\%, and so a high level of voltage unbalance in the network is not unreasonable). Without mitigation from the Soft Open Point, the three IMs can operate, but must be derated according to \eqref{e:derating_factor}. Key parameters for the case study are summarized in Table~\ref{t:case_b_parameters}. 

\begin{table}
\input{case_b_parameters}
\label{t:case_b_parameters}
\end{table}

The objective function, $\mathrm{OF}_2$, is to minimize the sum of the IM derating across the three IMs without causing heavy dc link ripple,
\begin{equation}
    \label{e:oF2}
    \mathrm{OF}_2 = C^{\mathrm{Derating}} + \beta \Pdcww
\end{equation}
where the derating cost $C^{\mathrm{Derating}}$ is from \eqref{e:cost_derating}, the dc link ripple power $\Pdcww$ is from \eqref{e:2w_p_bal}, and the weighting factor $\beta \ll 1$ is added as regularization to find operation which, where possible, avoids $\twoO$ ripple (as discussed in Section~\ref{sss:sop_zero_ripple}).

\begin{figure}\centering
\subfloat[Voltage unbalance at IM buses]{\includegraphics[width=0.24\textwidth]{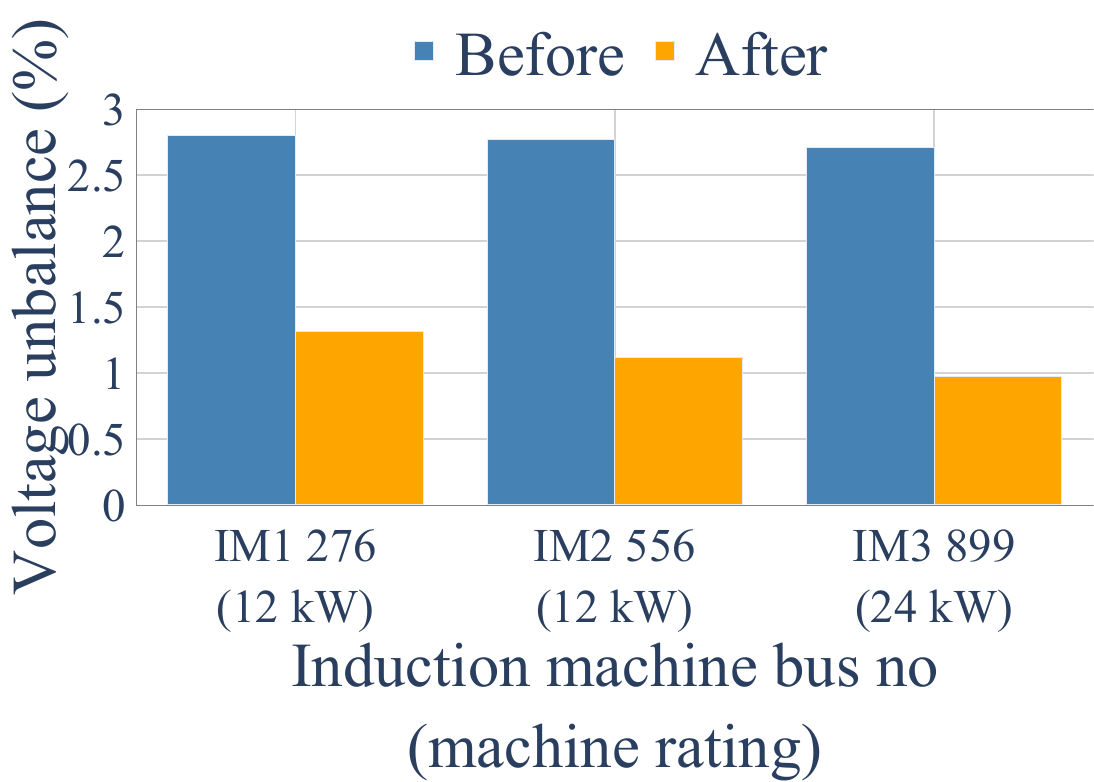}\label{f:pltSopBarUnbal}}
~
\subfloat[Required IM derating]{\includegraphics[width=0.24\textwidth]{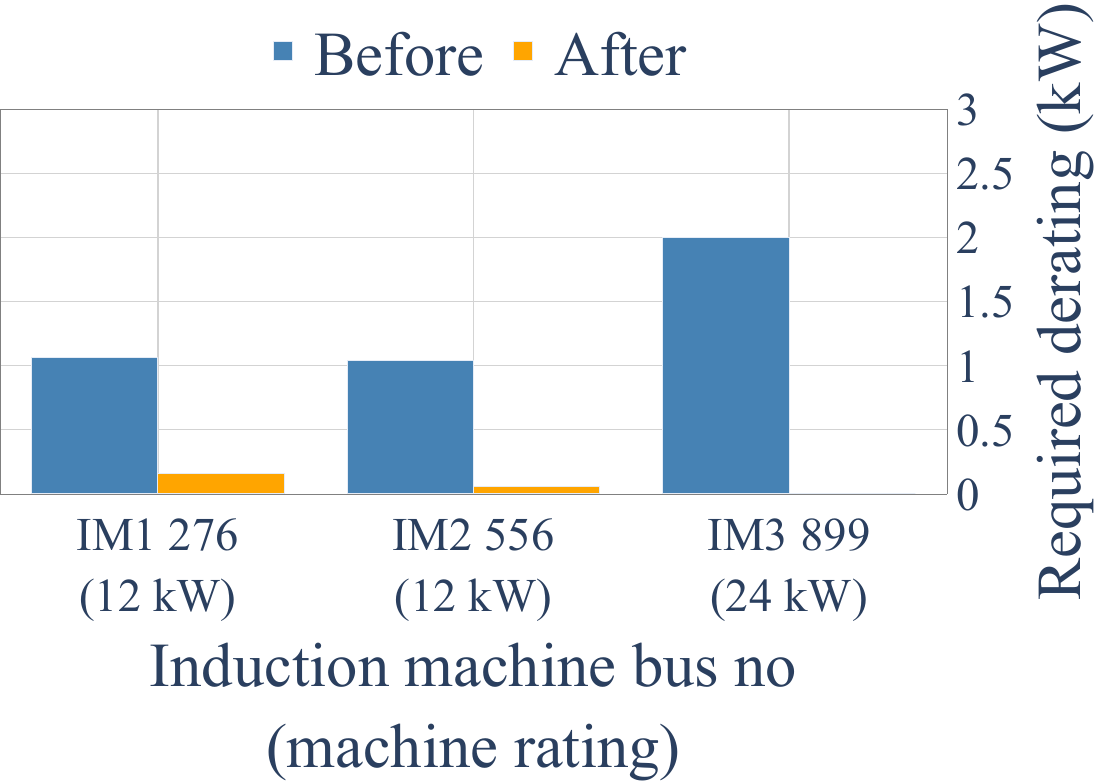}\label{f:pltSopBarCost}}
\caption{Case 2 results, showing the reduction in IM derating before and after the application of the Soft Open Point.}
\label{f:case_2_results}
\end{figure}

The results of this minimization for each of the three IMs are captured in Fig.~\ref{f:pltSopBarCost}. It can be seen that the derating is significantly reduced, with the total derating across the three machines reduced from 4.11~kW to 0.21~kW. This reduction of 94.9\% in IM derating is driven by a reduction in voltage unbalance at the three IMs, as shown in Fig.~\ref{f:pltSopBarUnbal}. These results demonstrate that the Soft Open Point can very effectively mitigate voltage unbalance issues and therefore the derating of these IMs.

The corresponding Soft Open Point current injections into feeders F1 and F2 at the optimal solution are presented in Table~\ref{t:case_b_sequence_currents}. It can be seen that injections into feeder F1 are dominated by a large negative sequence current. For this network, this is expected as the cable and transformer impedances are represented in terms of sequence components, so there is little coupling between negative sequence voltage and zero or positive sequence current injections. 

For feeder F2, there are no IMs or three-phase loads and voltage and current limits are met comfortably, so the Soft Open Point can inject current to mitigate the negative sequence component that is required to be injected into F1 (this can be seen as the phase angle between negative sequence currents is approximately 180 degrees). As a result, the numerical value of the ripple power $\Pdcww $ is negligible, taking value of just 0.52~W (compared to a F2 ripple power of 17.96~kW). It can be concluded that the generalized form \eqref{e:2w_p_bal} can conveniently account for $\twoO$ ripple for VSCs with more than four legs, enabling both three-phase and general multiterminal OPF problems to be considered using that same universal constraint.

\begin{table}
\input{case_b_sequence_currents}
\label{t:case_b_sequence_currents}
\end{table}

\subsection{Time domain analysis to validate \eqref{e:2w_p_bal}, \eqref{e:v_ripple}}\label{ss:res_ripple_validation}

To validate the impact of unbalance on low-frequency dc link ripple using the proposed bilinear form, time-domain simulations have been conducted using PLECS. We consider the ripple across Cases 3a--3f which qualitatively show the accuracy across a range of values of $I_j$, as shown in Table~\ref{tab:ripple_simulation_cases}.

\begin{table}
    \input{ripple_simulation_cases}
    \label{tab:ripple_simulation_cases}
\end{table}

\begin{figure}
 \centering
 \includegraphics[width=0.99\linewidth]{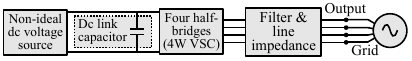}
 \caption{A 4W~VSC topology is used for time-domain simulations. The non-ideal dc voltage has a low-pass filter characteristic.}
\label{f:results_converter}
\end{figure}

Cases~3a--3f are studied for the Interconnected 4W~VSC topology shown in Fig.~\ref{f:results_converter}, with the main system parameters shown in Table~\ref{t:case_c_parameters}. At its output, the converter filter is inductive only; the line impedance is a resistance and inductance. The non-ideal dc voltage source consists of a series connected ideal dc voltage source, choke inductor and resistor, with the passive components selected to have the effect of a low-pass filter (7.5~Hz cutoff with damping factor $\zeta = 0.89$). Converter control for generating the unbalanced current injections is based on a bank of three proportional-resonant controllers \cite{teodorescu2006proportional} implemented in the `$\alpha \beta \gamma$' reference frame (via the Clarke transform). The output voltages of the converter are obtained through standard sinusoidal pulse width modulation (SPWM). 

\begin{table}
\input{case_c_parameters}
\label{t:case_c_parameters}
\end{table}

\subsubsection{Basic control validation}

\begin{figure}
 \centering
 \includegraphics[width=0.9\linewidth]{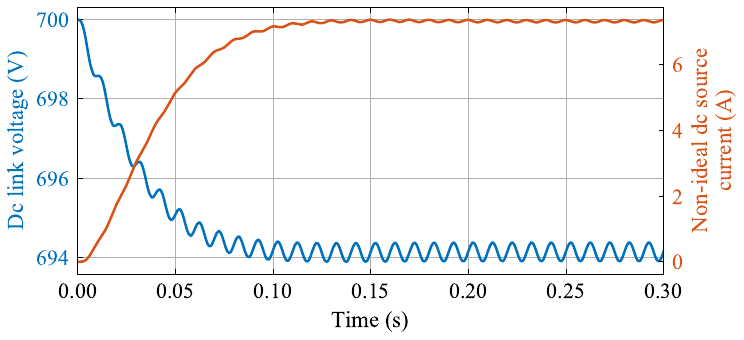}
 \caption{Voltage across the dc link capacitor and current injection from the non-ideal voltage source. The transient has decayed by the end of the time period shown (and therefore used subsequently to calculate the spectrum of the dc link ripple).}
\label{f:time_domain_dc_link}
\end{figure}

Fig.~\ref{f:time_domain_dc_link} plots the transient response of the dc link voltage and non-ideal dc voltage source's current for Case 3c. It can be seen that the non-zero injection of active power in the system leads to an initial reduction in dc link voltage as the capacitor initially discharges; the non-ideal dc voltage source then arrests the fall in the voltage by providing current (i.e., active power), with the dc link voltage stabilizing as the average dc active power across the capacitor becomes balanced.

\begin{figure}
\subfloat[Converter output (i.e., grid) voltages]{\includegraphics[width=0.48\textwidth]{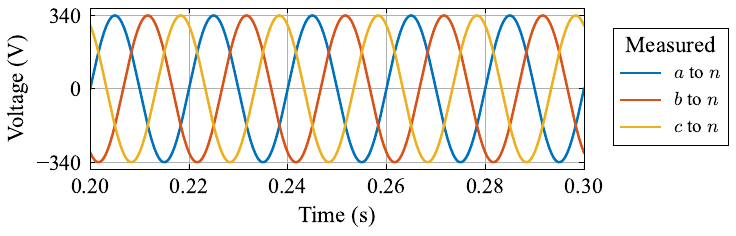}\label{f:time_domain_Vfeeder}}
~\\
\subfloat[Converter output currents]{\includegraphics[width=0.495\textwidth]{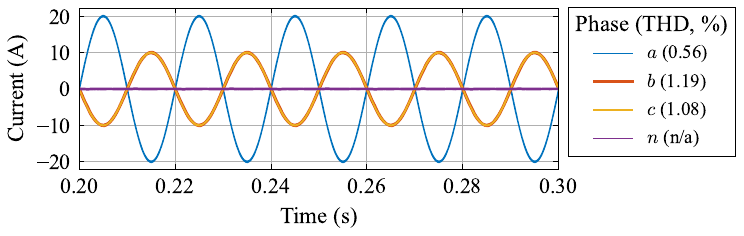}\label{f:time_domain_Ifeeder}}
\caption{The voltage and current at the converter output (i.e., grid-side of the line inductance) for Case 3c, showing stable operation with the controller correctly tracking current magnitude and phase, with low THD (b). }
\label{f:time_domain_feeder}
\end{figure}

The voltage and current at the output of the VSC system are shown in Fig.~\ref{f:time_domain_feeder}. Comparing these subfigures, it can be observed that the currents injected follow their reference correctly (Table~\ref{tab:ripple_simulation_cases}). The total harmonic distortion (THD), between 0.6\% and 1.2\%, is well within standard limits.

\begin{figure}
 \centering
 \includegraphics[width=1\linewidth]{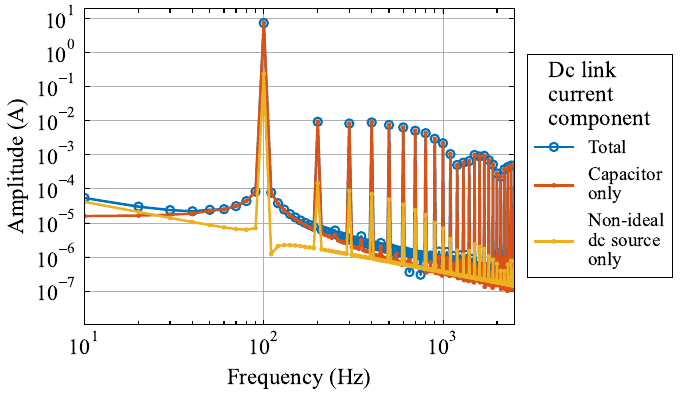}
 \caption{Spectrum of the dc link ripple up to order 50, capturing low-frequency ripple, for Case 3c (Phase A current injection into phase B and C), showing how the 100~Hz current ripple dominates the dc link current spectrum for the capacitor.}
\label{f:control_results}
\end{figure}

Fig.~\ref{f:control_results} shows the current spectrum of the dc link capacitor, non-ideal dc voltage source and total dc link current for Case 3c. It can be observed that there is a clear spike in at 100~Hz for the total dc link current (close to 10~A), with little ripple at other low frequencies (0.01~A or lower). At that frequency, the current ripple from the non-ideal voltage source (as can be seen in Fig.~\ref{f:time_domain_dc_link}) is also visible as a peak close to 0.1~A, but its magnitude is small compared to the capacitor's contribution.

\subsubsection{Verifying \eqref{e:2w_p_bal} and \eqref{e:v_ripple}}
To verify that the proposed bilinear form \eqref{e:2w_p_bal} and current ripple \eqref{e:v_ripple} effectively capture the dominant power and capacitor current ripple, the value of these quantities are collected in Table~\ref{tab:ripple_currents}. These illustrate that $\twoO$ power ripple is well-predicted by the proposed equations. Additionally, as the RMS values of both low-frequency current and power are very close to the value of the reported $\twoO$ values. It can be concluded that Cases 3a--3f demonstrate $\twoO$ ripple dominates the low-frequency RMS power and capacitor current ripple, and that the proposed formulation accurately determines the value of that ripple.

\begin{table}
    \input{ripple_currents}
    \label{tab:ripple_currents}
\end{table}

\section{Discussion}\label{s:discussion}

Traditionally, capacitor sizing in grid-connected VSCs has been driven by the need to store sufficient energy to avoid distortion and maintain stability after the largest expected power injection \cite{malesani1993ac,kolar2006analytical}. Looking to the future, however, several drivers suggest that dc link capacitors in VSCs will have less capacitance. First, capacitors have a large volume, so efforts to increase power density often consider total capacitance as a parameter to minimize \cite{blas2023sic}. Second, polypropylene film capacitors offer substantially increased reliability and lifetime as compared to aluminum electrolytic capacitors \cite{tayebi2018advanced,wang2014reliability}, but at the expense of lower energy density, providing further motivation for film capacitor-based dc links to minimize capacitance to maintain a reasonable VSC size. Finally, more powerful controllers combined with faster wide bandgap devices (e.g., Silicon Carbide or Gallium Nitride) enable much higher switching frequencies. These enable quicker disturbance response and reduced steady-state high-frequency power ripple compared to legacy VSCs, allowing smaller capacitors to achieve the same ripple performance under large unexpected active power injection.

Power ripple \eqref{e:2w_p_bal} holds for any grid-connected converter, but the impact on the capacitor loading \eqref{e:v_ripple} is not valid for other VSC topologies, such as multi-level converters, or VSCs with a split-phase capacitor (whose dc link capacitor also pass neutral current). Future work could explore interactions between power ripple and impacts on capacitor operations in those more complex topologies.

\section{Conclusions}\label{s:conclusions}

Power converters are endemic in power distribution systems, either as an interface for loads, generation, and energy storage, or increasingly as an approach for distribution network operators to control power flows. In this paper, we have incorporated the dominant $\twoO $ power ripple seen on VSCs operating in unbalanced networks as a factor to consider within distribution system OPF problems. Instantaneous power theory demonstrates the ripple is a bilinear function of the terminal voltages and currents, and so can readily be incorporated into existing OPF frameworks, whilst time-domain simulations verify the validity of the analysis. Proposed OPF problems demonstrates the impact of both hard constraints on ripple, and tradeoffs between ripple and competing network objectives. In back-to-back VSCs, it is shown that antiphase negative sequence injections into each connected feeder can enable unbalance mitigation with zero dc link ripple.

The move from balanced transmission-style OPF problems to unbalanced OPFs in distribution systems with active power converters will result in a need for improved understanding and appropriate consideration of operational constraints of these devices. This is of particular importance as increased communications and monitoring will enable converter control actions to be realized by network operators, with potential to increase operational efficiency and network capacity. It is concluded that the importance of consideration of physical constraints of different classes of power converters within unbalanced OPF problems will only increase.

\section*{Acknowledgment}

The authors are grateful for discussions with Frederik Geth (University of Queensland) on implementation of the proposed OPF.

\input{main.bbl}

\end{document}

%% file: defined_commands.tex
\newcommand{\tcell}[1]{\begin{tabular}[x]{@{}l@{}}#1\end{tabular}}

\newcommand{\twoO}{2\omega }

\newcommand{\zro}[1]{#1^{0}}
\newcommand{\pstv}[1]{#1^{+}}
\newcommand{\ngtv}[1]{#1^{-}}

\newcommand{\thVi}{\theta^{V}_{j}}
\newcommand{\thIi}{\theta^{I}_{j}}

\newcommand{\PdcDC}{P_{\mathrm{dc}}^{\mathrm{dc}}}
\newcommand{\Pdcww}{P_{\mathrm{dc}}^{2\omega}}
\newcommand{\VdcO}{V_{\mathrm{dc} }}
\newcommand{\IdcO}{I_{\mathrm{dc,src}} }

\newcommand{\Tfort}{T_{\mathrm{Fort.}}}

%% file: tables_final/table_OPF_IVR.tex
\centering
\caption{Proposed OPF Constraints and Objective Functions Considering $\twoO$ Ripple}
\setlength{\tabcolsep}{3.0pt}
\begin{tabular}{ll}
  \toprule
    VSC constraints \\
        \qquad current balance and limits & \eqref{e:iphase},\,\eqref{e:i_bal}\\
        \qquad power balance & \eqref{pdclink_pi}, \eqref{e:dc_p_bal} \\
        \qquad $\twoO$ dc link power and limit, & \eqref{e:2w_p_bal}, \eqref{e:2w_ripple_cons} \\
     Sensitive load constraints \\
        \qquad Negative sequence voltage limit & \eqref{e:Vneg_limit} \\
        \qquad Induction motor derating & \eqref{e:derating_factor}, \eqref{e:derating_func}

    \\
    Objective function & \eqref{e:oF1} or \eqref{e:oF2}
    \\
    \bottomrule
\end{tabular}

%% file: tables_final/case1.tex
\centering
\caption{20.7 kVA STATCOM operational constraints for Cases 1a--1d}
\begin{tabular}{llll}
\toprule
Case & \tcell{Phase leg\\capacity, A} & \tcell{Neutral leg\\capacity, A} & \tcell{$\twoO$ ripple power\\capacity, kW}\\
\midrule
1a & 30 & Unconstrained &  Unconstrained \\
1b & 30 & 0 &  Unconstrained \\
1c & 30 & Unconstrained &  0 \\
1d & 30 & 30 &  5 \\
\bottomrule
\end{tabular}

%% file: tables_final/case_b_parameters.tex
\centering
\caption{Soft Open Point and Network Parameters for Case 2}
\begin{tabular}{ll}
\toprule
Parameter & Value \\
\midrule
SOP rating (nominal total power) & 25~A per leg (34.5~kVA)\\
F1 negative sequence voltage (angle) & 0.0318~pu  (0$^{\circ}$)\\
Voltage upper bound & 1.10~pu \\
Voltage lower bound & 0.94~pu \\
IM1 power (lagging power factor) & 12 kW (0.85) \\
IM2 power (lagging power factor) & 12 kW (0.85) \\
IM3 power (lagging power factor) & 24 kW (0.85) \\
\bottomrule
\end{tabular}

%% file: tables_final/case_b_sequence_currents.tex
\centering
\caption{Soft Open Point Sequence Current Injections (solution of Case 2), with resulting dc link ripple $\Pdcww = 0.52$~W}
\begin{tabular}{llll}
\toprule
Current (feeder) & Solution value & Current (feeder) & Solution value \\
\midrule
$\zro{I}$ (F1) & 1.27$\angle$-132.0$^{\circ}$ & $\zro{I}$ (F2) & 0.07$\angle$-91.6$^{\circ}$\\
$\pstv{I}$ (F1) & 1.22$\angle$-95.6$^{\circ}$ & $\pstv{I}$ (F2) & 0.05$\angle$171.0$^{\circ}$\\
$\ngtv{I}$ (F1) & 24.94$\angle$-23.0$^{\circ}$ & $\ngtv{I}$ (F2) & 23.9$\angle$154.5$^{\circ}$\\

\bottomrule
\end{tabular}

%% file: tables_final/ripple_simulation_cases.tex
\centering
\caption{Case Studies for Simulating DC Link Ripple, $I_{\mathrm{Ref}}=14.14\angle 0$ A}
\begin{tabular}{lllll}
\toprule
Case ID & Description & $I_a/I_{\mathrm{Ref}}$ & $I_b/I_{\mathrm{Ref}}$ & $I_c/I_{\mathrm{Ref}}$  \\
\midrule
3a & Balanced $P$ & $1$ & $\alpha^{-1}$ & $\alpha^{-2}$ \\
3b & Phase $a,\,P$ only & $1$ & $0$ & $0$ \\
3c & Phase $a\,\,I$ to $b,\,c$ & $1$ & $-1/2$ & $-1/2 $ \\
3d & Negative sequence &  $1$ & $\alpha$ & $\alpha^2$ \\
3e & Balanced $Q$ & $\jmath$ & $\jmath \alpha^{-1}$ & $\jmath \alpha^{-2}$ \\
3f & Phase $a$, $Q$ only & $\jmath$ & $0$ & $0$ \\
\bottomrule
\end{tabular}

%% file: tables_final/case_c_parameters.tex
\centering
\caption{VSC time-domain simulation parameters (Cases 3a-3f)}
\begin{tabular}{llll}
\toprule
Parameter & Value & Parameter & Value \\
\midrule
RMS grid line voltage & 416~V & Grid frequency & 50~Hz \\
Switching frequency & 50~kHz & Peak current injection & 20~A \\
Dc link voltage & 700~V & Dc link capacitance & 50~mF \vspace{0.25em} \\
\tcell{Filter \& line\\inductance} & 5~mH & \tcell{Filter \& line\\resistance} & 1~m$\Omega$ \vspace{0.25em} \\
\tcell{Non-ideal dc\\source low-pass\\filter cutoff} & 7.5~Hz & \tcell{Non-ideal dc\\source low-pass\\filter damping factor} & 0.89 \\
\bottomrule
\end{tabular}

%% file: tables_final/ripple_currents.tex
\centering
\caption{Comparing proposed $\twoO$ power and capacitor current ripple, \eqref{e:2w_p_bal}, \eqref{e:v_ripple} respectively, against simulated values and low-frequency RMS values for Cases 3a-3f.}
\begin{tabular}{llllll}
\toprule
\multirow{3}*{\vspace{-0.4em} ID}& \multicolumn{2}{c}{$\twoO$ power ripple, kW} & \multicolumn{2}{c}{PLECS ripple $<$50$\omega$} & \multirow{3}*{\vspace{-0.4em} \tcell{$\twoO$ current,\\proposed\\\eqref{e:v_ripple},~A} } \\
\cmidrule(l{0.6em}r{0.9em}){2-3} \cmidrule(l{0.6em}r{0.9em}){4-5}
    & \tcell{Proposed,\\\eqref{e:2w_p_bal}} & \tcell{Simulated\\(PLECS)} & \tcell{Power,\\kW RMS} & \tcell{Current,\\A RMS} & \\
\midrule
3a & 0 & 6.68e-04 & 1.3e-03 & 1.9e-03 & 0 \\
3b & 3.458 & 3.466 & 3.466 & 4.979 & 4.940 \\
3c & 5.122 & 5.127 & 5.127 & 7.386 & 7.317 \\
3d & 10.200 & 10.207 & 10.208 & 14.582 & 14.571 \\
3e & 0 & 5.57e-03 & 5.7e-03 & 8.1e-03 & 0 \\
3f & 2.772 & 2.777 & 2.777 & 3.967 & 3.960 \\
\bottomrule
\end{tabular}